\documentclass[twocolumn,floatfix,superscriptaddress,amsmath,showpacs,showkeys,aps,prb]{revtex4}

\usepackage[final]{graphicx}
\usepackage{t1enc}
\usepackage{bm}

\begin{document}

\bibliographystyle{apsrev}

\title{\bf Long term ordering kinetics of the two dimensional q-state Potts model}

\author{Ezequiel E. Ferrero}
\email{ferrero@famaf.unc.edu.ar} \affiliation{Facultad de
Matem\'atica, Astronom\'{\i}a y F\'{\i}sica, Universidad Nacional
de C\'ordoba,
\\ Ciudad Universitaria, 5000 C\'ordoba, Argentina}
\altaffiliation{Members of CONICET, Argentina}

\author{Sergio A. Cannas}
\email{cannas@famaf.unc.edu.ar} \affiliation{Facultad de
Matem\'atica, Astronom\'{\i}a y F\'{\i}sica, Universidad Nacional
de C\'ordoba, \\ Ciudad Universitaria, 5000 C\'ordoba, Argentina}
\altaffiliation{Members of CONICET, Argentina}
\date{\today}

\begin{abstract}
We studied the non-equilibrium dynamics of the q-state Potts model
in the square lattice, after a quench to sub-critical
temperatures. By means of a continuous time Monte Carlo algorithm
(non-conserved order parameter dynamics) we analyzed the long term
behavior of the energy and  relaxation time for a wide range of
quench temperatures and system sizes. For $q>4$ we found the existence of
different dynamical regimes, according to quench
temperature range. At low (but finite) temperatures and very long times the
Lifshitz-Allen-Cahn
 domain growth behavior is interrupted with finite probability when the system stuck in highly symmetric
non-equilibrium metastable states, which induce activation in the
domain growth, in agreement with early predictions of Lifshitz
[JETP {\bf 42}, 1354 (1962)]. Moreover, if the temperature is very
low, the system always gets stuck at short times in a highly
disordered metastable states with finite life time, which have
been recently identified as glassy states. The finite size scaling properties
of the different relaxation times involved, as well as their temperature dependency are analyzed in detail.
\end{abstract}


\maketitle

\section{Introduction}

The problem of domain growth kinetics in systems with many
degenerate ground states had attracted a lot of attention in the
past\cite{Li1962,Sa1981,ViGr1987,GrAnSr1988,KuGuKa1987,SiMa1995}.
 In  early works Lifshitz\cite{Li1962} and then
Safran\cite{Sa1981} posed the discussion about the effect of
activated processes in the domain growth. They suggested that
$d$-dimensional, $q$-state degenerate models could become trapped
in local metastable
 states for $q \geq d+1$, which would then greatly slow down the relaxational
 kinetics. Their argument in two dimensions is that a honeycomb
 structure of hexagonal domains is stable under small distortions
 of the interfaces, because such distortions do not increase the
 interfacial free energy. Large distortions due to fluctuations  are then
 needed to move the interfaces and so domain growth becomes
 activated. The prototype of such a system is the q-state Potts
 model\cite{Wu1982}, whose Hamiltonian is given by

 \begin{equation}
H=- J~\sum_{nn}\delta (s_i,s_j) \qquad J>0 \label{Hamiltonian}
\end{equation}

\noindent where $s_i=1,2,\ldots,q$, $\delta (s_i,s_j)$ is the
Kronecker delta and the sum runs over all the pairs of
nearest-neighbor sites. Vi\~nals and Grant\cite{ViGr1987}
performed Monte Carlo (MC) simulations with Glauber dynamics for
 $q=8$  in square lattices of size up to
$N=128^2$ sites and time scales of the order $10^4$ Monte Carlo
Steps (MCS) (1 MCS is defined as  as a complete cycle of $N$ spin
update trials, according to a Metropolis MC algorithm). They found
evidences of metastable configurations composed by squares of
different colors (i.e., different values of $q$) that freeze the
dynamics a $T=0$. Moreover, they argue that those configurations
only effectively dominate at $T=0$ and growth at finite
temperatures is not limited by activated processes\cite{ViGr1987}.
Hence, for long enough times the average linear domain size should
follow the Lifshitz-Allen-Cahn (LAC) law $l(t) \sim t^{1/2}$.
Grest and coworkers\cite{GrAnSr1988} reported results using MC
simulations with $N=1000^2$ sites and values of $q$ up to
$q=64$ at very low temperatures, which are consistent with
LAC behavior for any value of $q$, at least for time scales up to
$10^4$ MCS. Analytical results on a coarse grained model also
confirm those results\cite{SiMa1995}. However, the time scales
considered in those works are very  short to exclude the existence
of activated process of the type predicted by
Lifshitz\cite{Li1962} and Safran\cite{Sa1981}. Moreover, recent
investigations\cite{Pe2003,OlPeTo2004,OlPeTo2004b} about the
pinning configurations found by  Vi\~nals and Grant\cite{ViGr1987}
showed that the system gets stuck in those states at times scales
longer than $10^4$ MCS; pinning also happens at low but finite
temperatures\cite{IbLoPe2006}, although in that case the
metastable states present a finite life time that increases
with\cite{IbFeCaLoPe2007} $q$. Those works also showed that the nature
of those highly disordered metastable states is more related to a
glassy one\cite{Pe2003,OlPeTo2004,OlPeTo2004b,IbLoPe2006} than to
the type of configurations predicted by Lifshitz\cite{Li1962} and
Safran\cite{Sa1981}. We will refer hereafter to those disordered
metastable states as the glassy ones.

 In this work we concentrated
mainly in the $q=9$ case in square lattices with $N=L\times L$
sites and periodic boundary conditions. Some complementary results
are also presented for other values of $q$. The implementation of
a continuous time MC algorithm or \emph{$n$-fold}
technique\cite{BoKaLe1975,No1995} allowed us to analyze a large
statistics on system sizes up to $L=500$ and time scales running
from $10^9$ to $10^{14}$ MCS. Some details of the continuous time
MC implementation are presented in Section \ref{methods}.

We analyzed the relaxation of the system after a quench from
infinite temperature down to subcritical temperatures $T$, for a
wide range of values of $T$ and $L$. We found the existence of
different dynamical regimes, according to the quench
temperature range. First of all there is some characteristic temperature
$T^* < T_c$, such that for $T^* <T < T_c$ simple coarsening
dominates the relaxation (that is, the domain growth follows the
LAC law even for very long times), except very close to $T_c$,
where evidences of nucleation relaxation mechanisms appear;
however, the last case will not be analyzed in this work and the
corresponding results will be presented in a forthcoming
publication.  For $T<T^*$ we found that, at long time scales
(i.e., much longer than those considered in previous works), the
LAC relaxation is interrupted when the system gets trapped in
highly symmetric metastable states with finite probability (i.e.,
for a large fraction of realizations of the thermal noise that do
not decrease as the system size increases). We found two different
types of those configurations: striped states and honeycomb like
structures; the latter are configurations composed by macroscopic
six-sided irregular polygonal domains of different colors. Striped
states are composed by two macroscopic ferromagnetic domains with
straight walls parallel to the coordinate axis and has been
previously observed in the Ising
model\cite{Li1999,SpKrRe2001a,SpKrRe2001b,SuSt2005,OlNeSiSt2006} ($q=2$)
at $T=0$. For $q\geq 3$ and $T=0$ the probability to reach a
striped state becomes zero in the thermodynamic limit\cite{SpKrRe2001b}; we found
that low but finite temperatures make that probability to become
finite.
The presence of honeycomb like structures in the q-state
Potts model, as far as we know, has not been previously reported
 and is in agreement with the Lifshitz's prediction\cite{Li1962}. When the system reaches either striped or
honeycomb like states the dynamics becomes activated.

Finally, we found a temperature $T_g\ll T_c$, such that
for $T <T_g $ the system always  gets stuck at
intermediate times in a glassy state of the type reported
previously\cite{Pe2003,OlPeTo2004,OlPeTo2004b,IbLoPe2006} for
other values of $q>4$.  We found that for $q=9$ those states
present a finite life time (i.e., independent of the system size)
with  a well defined free energy barrier associated to it.

We also
analyzed the scaling properties of the characteristic times associated
with the different relaxation processes, as well as the probability of
reaching a striped or honeycomb state for large values of $q$.

We verified that the whole relaxation scenario is qualitatively observed when
open(instead of periodic)
boundary conditions are used. All the
numerical results are presented in Section \ref{results} and some
discussion is presented in Section \ref{discussion}.

Besides its theoretical interest, the large-$q$ Potts model (or variations of it) is
used for simulating the dynamics of a
large variety of systems, such as soap froth\cite{GlAnGr1990,GlWe1992}, grain growth\cite{WeGl1992,ThAlGr2006} and
biological cells\cite{GrGl1992}. The present results help to establish
the conditions under which equilibrium can be actually reached, as well as the different
possible low temperature relaxation scenarios.

\section{Methods}
\label{methods}

We analyzed the time evolution under a type A dynamics
(non-conserved order parameter) of the system described by
Hamiltonian (\ref{Hamiltonian}), after a quench from infinite
temperature (i.e., a completely random initial configuration) down
to sub--critical temperatures. The Potts model undergoes a second
order phase transition for $q=2,3,4$ and a first order one for
$q>4$, where the critical temperature in the square lattice is known
exactly\cite{KiMiSh1954} for any value of $q$ and is given by
$T_c=1/\ln{\left(1+\sqrt{q}\right)}$ (hereafter we will use
natural units $k_B=J=1$). Most of our analysis were concentrated
in the $q=9$ case, for which $T_c=0.7213\ldots$, and some
complementary calculations were performed for $q=2,3,4,5,15$ and $30$.

 We were mainly interested in the late stages of the dynamics,
where large domains are formed. In that case, the computational
cost of a single spin flip dynamics (for instance, heat bath
algorithm) becomes very high, because the flipping probability of
spins inside the domains (which are the majority) becomes very
small. An efficient way to achieve longer time scales for
reasonable large system sizes is the usage of \emph{continuous
time} MC methods or \emph{$n$-fold} techniques \cite{BoKaLe1975}.
In these type of algorithms a flip occurs a each step and the time
such event would have elapsed in a single spin flip algorithm is
calculated from the associated flipping probability. Let us
briefly summarize the implementation of the algorithm for the
nearest-neighbors $q$-state Potts model. For a given  spin
configuration we will call ``potential spins'' to the $q-1$
possible states for each site in the lattice, different from the
present ones. All the potential spin  of the system are then
classified into lists, where members of a given list would produce
the same change in the energy of the system, if chosen to replace
the old spin state in the corresponding site. For a single flip
there exist only $9$ possibilities ($\frac{\Delta E}{J} =
-4,-3,-2,-1,0,1,2,3,4$), so we have $9$ classes for any value of
$q\geq 3$. In the Metropolis algorithm the probability of a
potential spin belonging to the class $l$ ($l=1,\ldots,9$) to be
effectively flipped  is:

\begin{equation} \label{p_l}
p_l = \frac{n_l}{N(q-1)}\; {\rm min}\left[ 1, \exp{
\left(-\frac{\Delta E_l}{k_B T}\right)} \right]
\end{equation}

\noindent where $n_l$ is the number of spins in class $l$. The
total probability of any flipping event occurring in a given step
is

\begin{equation} \label{Q_M}
Q=\sum_l p_l
\end{equation}

\noindent In the present algorithm, at each step a class is sorted
with probability (\ref{p_l}) and a potential spin is sorted with
equal probability among all the members in the class. After
updating the corresponding spin and the lists, the time step is
incremented by an amount

\begin{equation}
\Delta t=\frac{-ln~ r}{N\, Q}
\end{equation}

\noindent where $r$ is a random number uniformly distributed
between zero and one and the time step $\Delta t$ is measured in
MCS. The details of the algorithm can be seen in
Ref.\cite{BoKaLe1975}. The implementation of this algorithm
allowed us to perform simulations for time scales running from
$10^9$ MCS (for sizes up to $L=500$) to $10^{14}$ MCS (for sizes
up to $L=100$). In order to check the algorithm we also repeated
several of the simulations using a single spin-flip algorithm
(heat bath) for $L=200$ and time scales up to $10^6$ MCS. The
results were identical.

Our analysis of the dynamics was mainly focused on the behavior of
two quantities: the average energy per spin $e(t) \equiv \left<
H(t) \right>/N$ as a function of $t$ (the average was taken over
different initial configurations and different realizations of the
thermal noise) and the {\it equilibration time} $\tau$; the last
quantity was defined as the time at which the instantaneous energy
falls below an equilibration threshold. Such threshold was set as
the equilibrium energy at the corresponding temperature plus one
standard deviation, where those quantities were first calculated
by running a  set of simulations starting from the ordered state
and letting the system to equilibrate. We calculated the
probability distribution (normalized histogram) $P(\tau)$ for
different values of $T$ and $L$.

\section{Results}

\label{results}

\begin{figure}
\begin{center}
\includegraphics[scale=0.3]{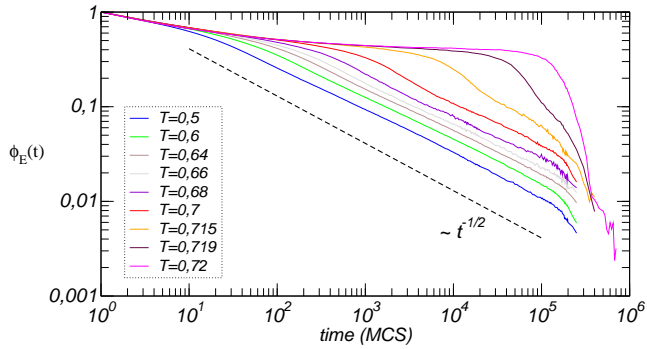}
\caption{\label{fig1} (Color on-line) Relaxation function
$\phi_E(t)$ for $L=300$, $q=9$ and temperatures ranging from
$T=0.5$ (bottom) to $T=0.72$ (top).}
\end{center}
\end{figure}

In order to compare the behavior of the average energy per spin
$e(t) =\left<H \right>/N$ for  different temperatures, we first
introduce the relaxation function (or normalized excess of energy)

 \begin{equation}
\phi_E(t) \equiv \frac{e(t) -e(\infty)}{e(0)-e(\infty)}
 \end{equation}

\noindent where $e(\infty)$ is the equilibrium energy.  In
Fig.\ref{fig1} we show the typical behavior of $\phi_E(t)$ for
$L=300$ and temperatures between $0.72$ and $0.5$
($T_c=0.7213\ldots$). For temperatures close enough to $T_c$ ($
0.715 < T < T_c$) we see that the system clearly stuck in a high energy
metaestable state. Close examination of different quantities in
the metastable state show that this corresponds to a disordered
(i.e., paramagnetic) one and hence it is directly related to the
first order nature of the transition. Moreover, we found evidences
that in this regime relaxation is dominated by nucleation
mechanisms, but the details of that analysis will be presented in
a forthcoming publication. For temperatures $T < 0.715$ we see
that the metastable plateau disappears and the relaxation function
decays (after a short transient) for all temperatures as
$\phi_E(t) \sim t^{-1/2}$. Since the excess of energy respect to
the equilibrium state in a domain growth process is given by the
average energy of the domain walls, a simple calculation shows
that $\phi_E(t) \sim 1/l(t)$, $l(t)$ being the average linear
domain size. Hence, the behavior of Fig.\ref{fig1} is consistent
with the LAC law. As we will show later, the finite size scaling
properties of the average typical equilibration time in this
temperature range are also consistent with the LAC law.

\begin{figure}
\begin{center}
\includegraphics[scale=0.3]{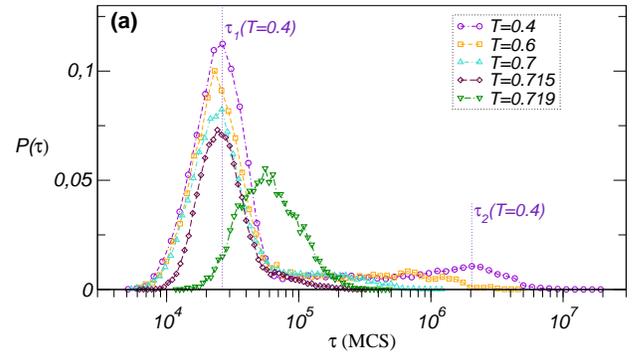}

\vspace{1.2cm}

\includegraphics[scale=0.3]{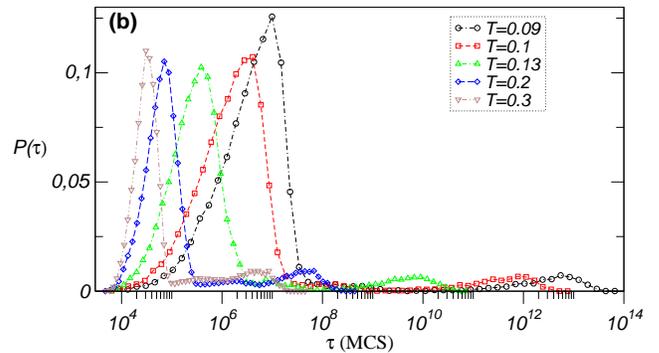}
\caption{\label{fig2} (Color on-line) Equilibration time
probability distribution $P(\tau)$ for $L=100$ and $q=9$. (a)
Temperatures ranging from $T=0.4$ (top) to $T=0.719$ (bottom);
(b) temperatures ranging from $T=0.3$ (left) to $T=0.09$ (right).}
\end{center}
\end{figure}

\subsection{Relaxation at intermediate temperatures and blocked states: characterization and scaling}

\begin{figure}
\begin{center}
\includegraphics[scale=0.32]{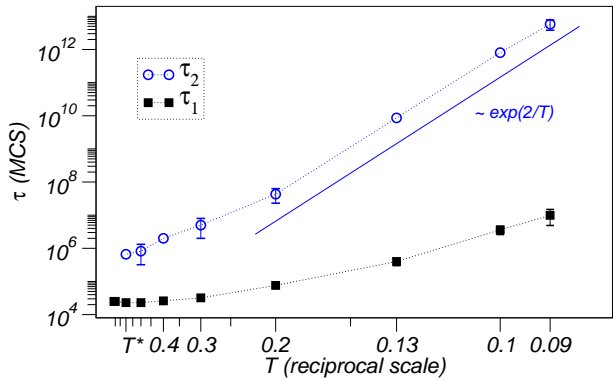}
\caption{\label{fig3} (Color on-line) Characteristic relaxation
times $\tau_1$ and $\tau_2$ {\it vs.} $1/T$ for $L=100$ and $q=9$;
the same qualitative behaviors are observed for other system
sizes. The continuous lines are a guide to the eye.}
\end{center}
\end{figure}

In Fig.\ref{fig2} we see the typical behavior of $P(\tau)$ for an
intermediate  size ($L=100$) and different temperature ranges. The
different dynamical regimes can be already appreciated in this
figure. Close enough to $T_c$ ($T=0.719$ in Fig.\ref{fig2}a)
$P(\tau)$ exhibit a well defined peak centered at a characteristic
time $\tau_{nucl} \sim 10^5$ MCS, which is associated to a
nucleation based relaxation mechanism already mentioned. As the
temperature decreases below some temperature $0.715 < T_{n} <
T_c$, this peak is suddenly replaced by another one centered at a
characteristic value $\tau_1$, which is about one order of
magnitude smaller of $\tau_{nucl}$ and remains almost independent
of the temperature in the range $0.3 < T < 0.715$; in the
temperature range $0 < T \leq 0.2$ (see Fig.\ref{fig2}b), $\tau_1$
exhibit a strong temperature dependency. For temperatures smaller
than (but close to) $T_n$ (see Fig.\ref{fig2}a), $P(\tau)$ develops
a long right tail; for temperatures $T< T^* \approx 0.6$ the tail
becomes a distinct peak centered at a new characteristic time
$\tau_2$, which increases exponentially as the temperature
decreases. This behavior indicates the existence of two distinct
phenomena affecting the relaxation at different time scales, where
$T^*$ acts as a reference temperature signaling the time scales
separation crossover point. The temperature behavior of $\tau_1$
and $\tau_2$ is summarized in the Arrhenius plot of Fig.\ref{fig3}
.

\begin{figure}
\begin{center}
\includegraphics[scale=0.35,angle=-90]{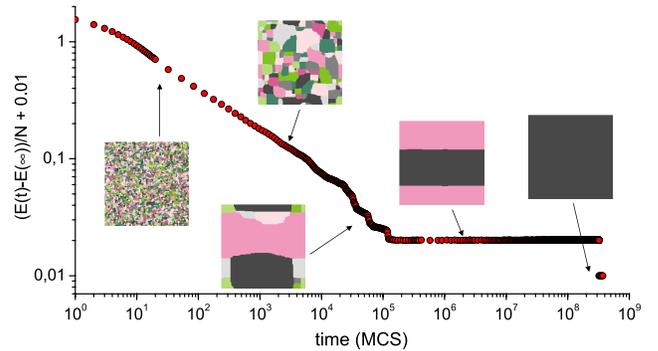}
\caption{\label{fig4} (Color on-line) Energy per spin as a
function of time and typical spin configurations in one
realization of the stochastic noise, when the system gets stuck in a
striped configuration ($L=200$, $q=9$ and $T=0.2$). The different
colors codify different spin values $s_i=1,\ldots,9$}
\end{center}
\end{figure}

We will show  that $\tau_1$ is associated with simple coarsening
processes that follow LAC law for all times, while $\tau_2$ is
associated to processes in which the system gets stuck in striped
metastable states, composed by two ferromagnetic states whose
walls are parallel to coordinate axis, as shown in the example of
Fig.\ref{fig4}. Those type of metastable states have been already
observed in the two dimensional Ising model ($q=2$) at zero
temperature, where they become
frozen\cite{Li1999,SpKrRe2001a,SpKrRe2001b,SuSt2005,OlNeSiSt2006}. At
finite temperature, striped states perform a random parallel
movement in the direction perpendicular to the walls. Hence, in a
finite system those states relax to equilibrium when  both walls
collapse. Spirin, Kaprivsky and Redner\cite{SpKrRe2001a} showed
that the basic mechanism for the parallel movement of a straight
domain wall is the creation of a ``dent'', that is, the flip of
one of the spins adjacent to the wall. Since after  flipping
 the spin its neighbors can flip without energy cost, the energy
barrier for the creation of a dent is $2$ (in units of the
coupling
constant $J$)  for the $q=2$ Potts model (or $4$ for the Ising
model). For $q
>2$ the energy cost of any other movement (including a flipping to
a third color different from those of the domains) is larger.
Hence, once the striped state is reached, the time needed to relax
should be basically independent of $q$ and this is consistent with
the Arrhenius behavior $\tau_2 \sim e^{2/T}$ observed in
Fig.\ref{fig3}.

From Fig.\ref{fig3} we can notice also that, for a wide range of
temperatures $T< T_n$ (approximately down to $T \approx 0.2$)
$\tau_1$ remains almost independent of $T$, consistently with a
simple coarsening behavior; at lower temperatures we see a
crossover into an activated behavior, that will be analyzed later.

\begin{figure}
\begin{center}
\includegraphics[scale=0.3]{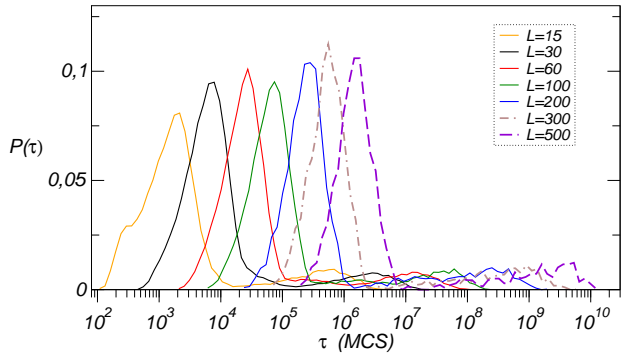}
\caption{\label{fig5} (Color on-line) Equilibration time
probability distribution $P(\tau)$ for $T=0.2$, $q=9$ and system
sizes ranging from $L=15$ (left) to $L=500$ (right).}
\end{center}
\end{figure}

A deeper understanding of the mechanisms involved in the
relaxation can be obtained from the finite size scaling of the
different quantities involved. In Fig.\ref{fig5} we show the
typical behavior of $P(\tau)$ for different system sizes at a
fixed temperature $T=0.2$. The first thing we note is that the
two-peak structure remains in the large $L$ limit.  Moreover, the
ratio between the areas below both peaks becomes constant in such
limit. The same property is observed for
temperatures up to $T^*$. We will analyze this in more detail at the
end of this section. Let us now consider the finite
size scaling of the relaxation times.

\begin{figure}
\begin{center}
\includegraphics[scale=0.3]{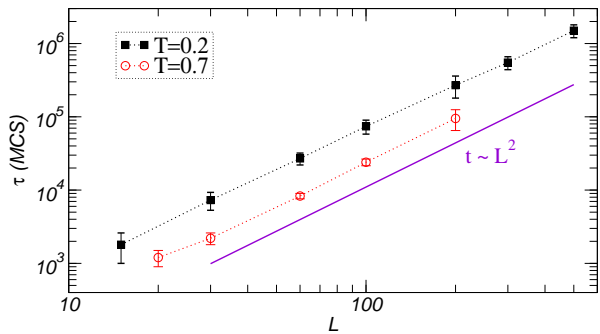}
\caption{\label{fig6} (Color on-line) Characteristic relaxation
time $\tau_1$ {\it vs} $L$ for $q=9$ and different temperatures; dotted lines are a guide to the eye.}
\end{center}
\end{figure}

From Fig.\ref{fig6} we see that $\tau_1 \sim L^2$ for a wide range
of temperatures, both above and below $T^*$. This is also
consistent with a simple coarsening process, in which
equilibration will be attained once $l(\tau) \sim \tau^{1/2} \sim
L$.

\begin{figure}
\begin{center}
\includegraphics[scale=0.3]{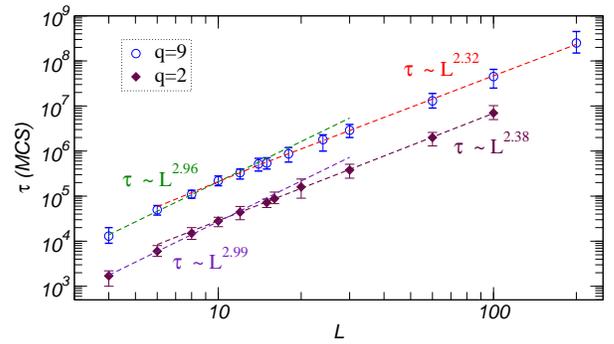}
\caption{\label{fig7} (Color on-line) Characteristic relaxation
time $\tau_2$ {\it vs} $L$ for $T=0.2$ and different values of
$q$. The dashed lines correspond to linear fittings.}
\end{center}
\end{figure}

\begin{figure}
\begin{center}
\includegraphics[scale=0.3]{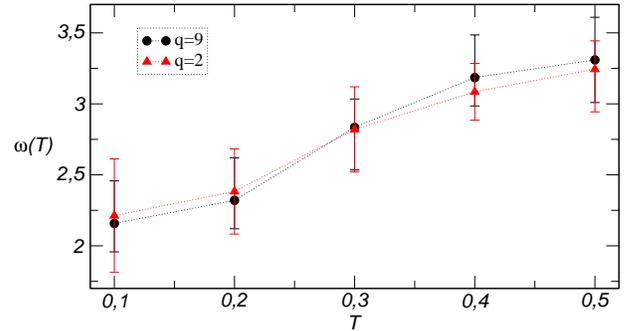}
\caption{\label{fig8} (Color on-line) Finite size scaling exponent
$\omega$ of the characteristic relaxation time $\tau_2$ as a
function of $T$ for different values of $q$; dotted lines are a guide to the eye.}
\end{center}
\end{figure}

Let us now analyze the finite size scaling of $\tau_2$. Spirin,
Kaprivsky and Redner\cite{SpKrRe2001a} suggested that, at low
enough temperatures, the movement of a flat interface will be
dominated by processes involving a single dent creation; once the
dent is created it performs a random walk, until either the dent
disappears or it covers the whole line, where the typical time
needed for the last event scales as\cite{SpKrRe2001a} $L$. This
mechanism leads to a random walk movement of both interfaces, so
there must be typically $L^2$ such hopping events for the
interfaces to meet and therefore the relaxation time should scale
as\cite{SpKrRe2001a} $L^3$. However, this argument only works for
small system sizes. Once a dent is created, the probability of
creation of new dents along the interface, before the dent covers
the line, increases with the system size; therefore, the typical
time for a one-site hoping event of an entire interface should
increase slower than linearly with $L$ and $\tau_2$ slower than $L^3$. This can be appreciated in
the clear crossover from $\tau_2\sim L^3$ to $\tau_2\sim L^\omega$
with $\omega <3$ around $L=15$,  observed  in Fig.\ref{fig7}, both
for $q=2$ and $q=9$ (the same effect is observed for any temperature $T \leq 0.2$).
Striped states appear for temperatures up to $T^*$, but the
the walls movements are no longer dominated by one-site hopping events  for $T> 0.2$;
instead of that, a direct inspection of the spin configuration
during relaxation shows that, at temperatures close to $T^*$ the movements of the domain
walls resemble (for large system sizes) that of  elastic lines
subjected to a random noise. Hence, the
temperature dependence of $\tau_2$ departs from the $e^{2/T}$
behavior, as can be seen from Fig.\ref{fig3}.
However, the finite size scaling $\tau_2\sim L^\omega$ still holds
for temperatures up to $T^*$, where the exponent $\omega$ displays
a marked increase with the temperature, reaching values slightly
larger than $3$ as $T$ approaches $T^*$ (see Fig.\ref{fig8}).
Those values of the exponent can be understood through the following argument.  Suppose that each line behaves
as a chain of $L$ unit masses joined by springs, constrained to move along the
direction perpendicular to the wall and subjected to independent
white noise. By solving the corresponding Langevin equations in the overdamped limit,
a simple calculation shows that the  distance
between the centers of mass of both chains performs a Brownian
motion with an effective diffusion coefficient that scales as $D
\sim L^{-1}$. Since the distance between walls is of the order of
$L$, this implies that the typical time needed for the walls to
encounter should scale approximately as $L^3$. For $q=2$ Lipowski\cite{Li1999} has shown
that this scaling holds even for relatively large values of the
temperature $T/T_c(2) \approx 0.8$. The results of Fig.\ref{fig8}
suggest that the scaling properties of $\tau_2$ are independent of
$q$, showing that large degeneracies in the ground state have no
influence in this relaxation process.

\begin{figure}
\begin{center}
\includegraphics[scale=0.3,angle=-90]{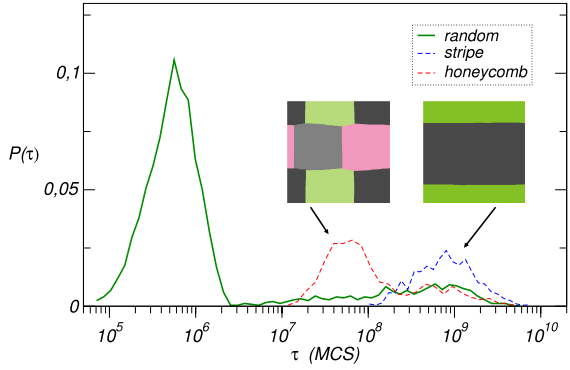}
\caption{\label{fig9} (Color on-line) Equilibration time
probability distribution $P(\tau)$ starting from different initial
configurations (arbitrary normalization)  for $L=300$, $T=0.2$.
The inset images show typical blocked spin configurations at the
corresponding times.}
\end{center}
\end{figure}

Let us return to the equilibration time probability distribution
$P(\tau)$. Another salient feature of this distribution for
temperatures $T<T^*$ is that the right peak broadens for large
system sizes. To show this we redraw $P(\tau)$ for $L=300$ and
$T=0.2$ in Fig.\ref{fig9} (full line). A careful inspection of individual
processes shows that, such broad peak is actually associated with
two different types of metastable configurations: the striped ones
already described and honeycomb like structures; the latter are
composed macroscopic six-sided irregular polygonal domains of
different colors (see inset of Fig.\ref{fig9}), where the angles
between domain walls at the three-fold edges fluctuate around
$120^o$. Those states are in agreement with Lifshitz
prediction\cite{Li1962} for $q \geq 3$ and we shall call them
Lifshitz states. By a  calculation of the equilibration time
starting directly from the Lifshitz and striped states, we
verified that the broad peak of $P(\tau)$ is actually a
superposition of two peaks, each one  with its own distinctive
maximum at characteristic times $\tau_2$, for the striped states,
and $\tau_3$ for the Lifshitz ones (see Fig.\ref{fig9}). Lifshitz
states are only detectable for system sizes $L\geq 100$. Actually,
isolated three-fold vertex between flat domain walls of the type predicted
by Lifshitz\cite{Li1962} also appear for smaller system sizes, but
complete honeycomb-like structures can be stabilized during
detectable time scales (i.e., larger than the characteristic
coarsening time scales) only for large enough system sizes.
To determine the
scaling properties of $\tau_3$, we calculated the escape time
probability distribution starting from the closest configuration
to a  Lifshitz state, that is, from an almost perfect four-colored
honeycomb configuration (commensurability with the system size
does not always allow a perfect honeycomb structure) for different
values of $L$ and $T$; we show an example in Fig.\ref{fig9}. We verified that the system quickly relaxes from that configuration
 into a Lifshitz state, from which it
can either relax directly to the equilibrium state or pass first
to a striped state, giving rise to a second peak in the
corresponding probability distribution (see Fig.\ref{fig9}). For
completeness, we also calculated the escape time probability
distribution starting from a perfect two-domains striped state;
the result is also shown in Fig.\ref{fig9}.

\begin{figure}
\begin{center}
\includegraphics[scale=0.3]{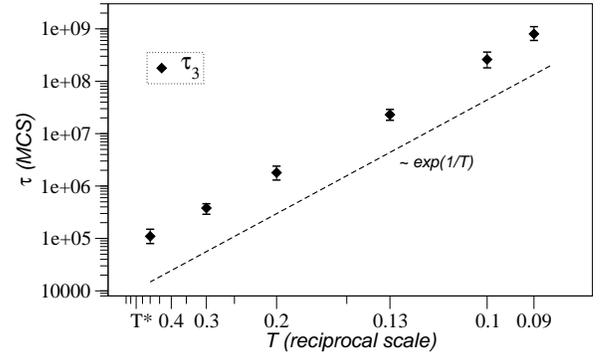}
\caption{\label{fig10}  Characteristic relaxation time $\tau_3$
{\it vs} temperature for $q=9$ and $L=100$; the continuous line is a guide to the eye.}
\end{center}
\end{figure}

\begin{figure}
\begin{center}
\includegraphics[scale=0.6]{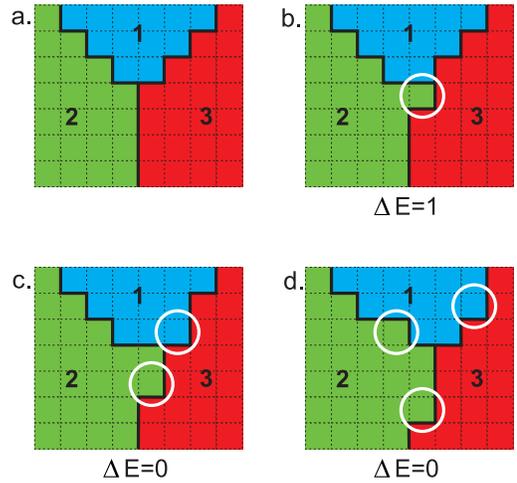}
\caption{\label{fig11} (Color on-line) Basic relaxation mechanism  of a Lifshitz state at low temperatures.
The circle in (b) marks the creation of a dent in the vertex of (a), by flipping a spin from 2 $\rightarrow 3$.
Circles in (c) and (d) exemplify spins that can flip without energy cost along the three converging walls.
The whole process may lead to the hopping of the whole structure (vertex plus walls) and ultimately to the collapse of two vertex.}
\end{center}
\end{figure}

The temperature dependency of $\tau_3$ is shown in
Fig.\ref{fig10}, where we see that it displays a clear Arrhenius
behavior with an activation barrier of height one, which is the
minimum possible energy barrier associated with a single spin
flip. This can be understood if we analyze the basic mechanisms
behind the relaxation from Lifshitz states. We observed that
Lifshitz states relax when two vertex of an hexagon edge collapse.
A vertex movement, with the consequent displacement of the converging walls,
occurs through a series of random hopping events. In Fig.\ref{fig11}
we show an example of the hopping of a vertex one site to the right;
one-site hopping events  in the other directions follow a similar process with the same energy cost.
 The movement of a vertex starts with the creation of a dent, by
flipping  one of the spins located at the neighbor sites of the
vertex, as depicted in Fig.\ref{fig11}b. This movement has an
energy cost of one unit. Once the dent is created, the neighbor spins
at the three converging walls are free to flip without energy cost (see Figs.\ref{fig11}c and \ref{fig11}d),
generating a diffusive motion of the dent along the three lines, and may
lead to the displacement of the whole lines. This
hopping movement of the vertex
ultimately leads to the collapse of two of them and the consequent
disappearance of the Lifshitz state. The whole
mechanism is completely similar to that described by Spirin,
Kaprivsky and Redner\cite{SpKrRe2001a} in the case of a flat wall
between two striped domains, except that the creation of a dent
adjacent to a vertex has an energy cost of just one energy unit (instead of
2,
as in the case of a dent in a flat interface), which explains the
behavior of Fig.\ref{fig10}. Since hexagonal domains in a Lifshitz state are macroscopic, the same finite size scaling
arguments used by Spirin, Kaprivsky and Redner\cite{SpKrRe2001a}
for the relaxation time apply in this case. Hence, one expect $\tau_3 \sim L^\omega\;
e^{1/T}$. For the system sizes available, we verified this scaling
at low temperatures with an exponent $\omega \approx 3$, but we
would expect this value to be reduced for larger system sizes, as
in the case of  striped states ($\tau_2$).

\subsection{Probability of blocked states}

We analized the probability
$P_b(q)$ of getting stuck in a
blocked state. We defined blocked states as those characterized by flat walls between domains.
For $q\geq 3$ this includes Lifshitz and striped states. For $q=2$
the system can also get trapped in another type of blocked states,
characterized by diagonal stripes, whose interfaces  fluctuate without
energy cost\cite{SpKrRe2001b}; we shall call them diagonal states. For $q\geq
3$ we did not observed diagonal states at finite temperatures.
Although their presence for $q\geq
3$ with low probability cannot be excluded,
probably they are replaced at finite temperature by the Lifshitz
states.

From the previous calculations of $P(\tau)$ we could estimate
$P_b(q)$ by defining for (for every value of $T$ and $L$) a
threshold value $\tau_t(T,L)$, such that a single realization with
$\tau > \tau_t$ is attributed to the presence of a blocked state;
$\tau_t$ can be estimated as the first minimum of $P(\tau)$
located above $\tau_1$ (see for instance Fig.\ref{fig5}). This
procedure reduces the calculation of $P_b(q)$ to a binomial
experiment. hence, a simple calculation shows that a sample size
of $2000$ runs is enough to guarantee a statistical error smaller
than $1\%$ in all cases, thus saving a lot of CPU time.

\begin{figure}
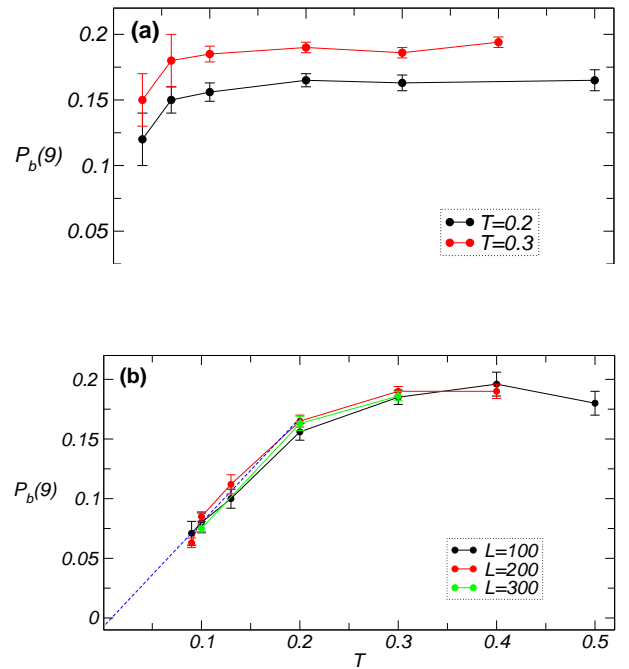

\begin{center}
\includegraphics[scale=0.31]{fig12A.eps}
\vspace{0.5cm}
\includegraphics[scale=0.29]{fig12B.eps}
\caption{\label{fig12} (Color online) Probability of getting stuck in a blocked state $P_b(q)$
for $q=9$: (a)  as a function of $L$ for different temperatures; (b) as a function of $T$ for different sizes.
The dashed line corresponds to a linear fitting of the points for $L=200$, giving an extrapolated value of $0.008\pm 0.01$ at $T=0$.}
\end{center}
\end{figure}

In Fig.\ref{fig12} we show the results for $q=9$. The main source of error in this calculation
is the choice of $\tau_t$, which is not always evident, due to large fluctuations
in
the histograms for small sizes and very low temperatures;
the error bars in Fig.\ref{fig12} were estimated by varying $\tau_t$. From
Fig.\ref{fig12}a we see that, at $T > 0$, $P_b(9)$ saturates in a finite
value for $L \geq 100$, indicating a finite probability in the limit $L\rightarrow\infty$. In Fig.\ref{fig12}b we show the
temperature dependency of the saturation value. We see that
$P_b(9)$ goes to zero as $T\rightarrow 0$, consistently with the
results of Spirin, Kaprivsky and Redner\cite{SpKrRe2001b}.

\begin{figure}
\begin{center}
\includegraphics[scale=0.3]{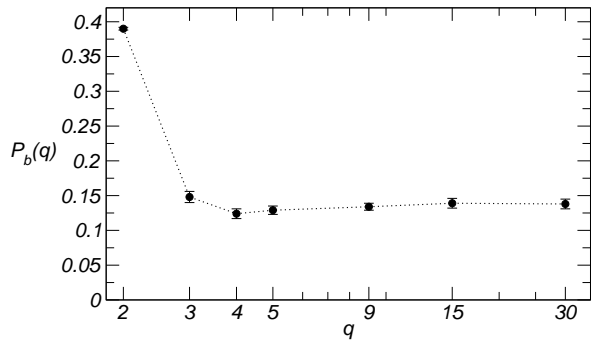}
\caption{\label{fig13} Probability of getting stuck in a blocked state $P_b(q)$ for
$L=200$ and $T=0.15$.}
\end{center}
\end{figure}

Next we analyzed the probability $P_b(q)$ as a function of $q$. The results are
shown in Fig.\ref{fig13} for $T=0.15$ and values of $q$ ranging from $q=2$ up to
$q=30$ .
For $q=2$ and $T=0$ the probability of reaching a striped state is\cite{SpKrRe2001a,FiHo2005} $1/3$,
while the probability of reaching a diagonal state is\cite{SpKrRe2001b} $\approx
0.04$. At $T=0.15$ we found the values $\approx 0.345$ and $\approx 0.045$
respectively, giving rise to the value $P_b(2)\approx 0.39$. The
differences with the $T=0$ values are consistent with the
enhancement of the probability at finite temperature, already
observed for $q=9$.

For $q\geq 3$ the probability $P_b(q)$ falls down to a temperature
dependent finite value, that is almost independent of $q$ and smaller than
half of $P_b(2)$.

It is worth noting that Spirin, Kaprivsky and
Redner\cite{SpKrRe2001b} reported another type of blocked states
for $q=3$, characterized by both straight walls and diagonal
walls, the latter fluctuating without energy cost; they call these
states ``blinkers''. We did not observe blinkers at finite
temperature, at least for periodic boundary conditions. Although their existence with low probability cannot
be excluded, probably they decay into Lifshitz states in time
scales smaller than the characteristic Lifshitz relaxation times.

\subsection{Low temperatures relaxation: glassy states}

\begin{figure}
\begin{center}
\includegraphics[scale=0.3]{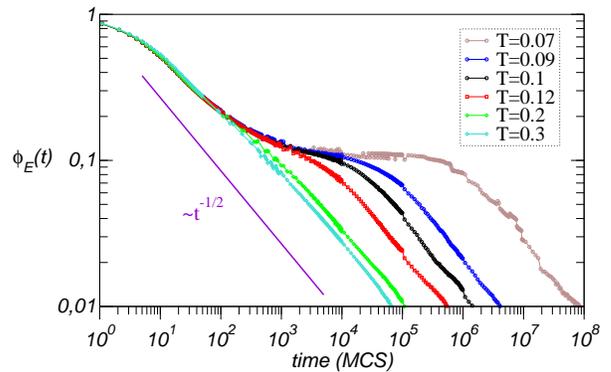}
\caption{\label{fig14} (Color on-line) Relaxation function for $q=9$, $L=200$ and different temperatures around $T_g$. Temperatures increase from right to left.}
\end{center}
\end{figure}

\begin{figure}
\begin{center}
\includegraphics[scale=0.35,angle=-90]{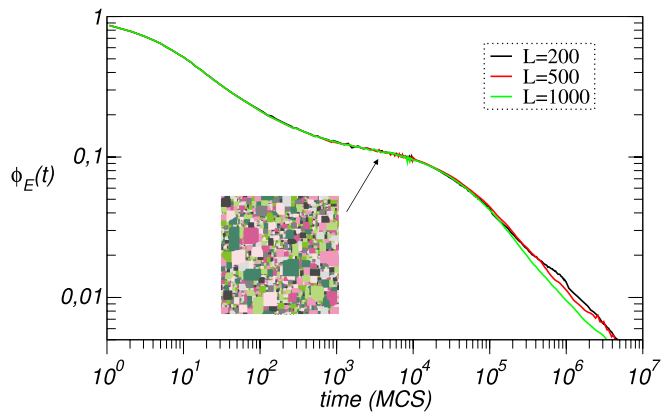}
\caption{\label{fig15} (Color on-line) Relaxation function for $q=9$, $T=0.1$ and different values of $L$.
The inset shows a typical configuration of the glassy state associated with the plateau.}
\end{center}
\end{figure}

 Let us now
analyze the coarsening at very low temperatures. The increase of $\tau_1$ for
temperatures $T< 0.2$ observed in Fig.\ref{fig3} indicates that the normal coarsening is affected by some kind of
 activated process. The increase in this relaxation time is associated with the
plateau displayed by the relaxation function in Fig.\ref{fig14}. We
found that this plateau appears below some characteristic temperature $0.1 < T_g
< 0.2$ for $q=9$. This plateau corresponds to a disordered metastable
state characterized by almost square--shaped domains with a wide
distribution of sizes (see inset in
Fig.\ref{fig15}). That type of metastable state was
previously reported for $q=7$ and it was identified as a glassy
one\cite{Pe2003,OlPeTo2004,OlPeTo2004b}.These states are only present for\cite{IbLoPe2006} $q>4$. We verified that
for $q=9$ the normal coarsening is always interrupted for $T<T_g$ and the system gets stuck in one of those glassy
states, from which it relaxes through a complex sequence of
activated jumps. This explains the exponential increase of
$\tau_1$ observed in Fig.\ref{fig3} for  $T< 0.2$. Once the
system relaxes from the glassy state, it can be either directly equilibrate
or decay first in a blocked state.

In Fig.\ref{fig15} we show the typical behavior of relaxation function for $q=9$ at a fixed temperature $T<T_g$
and different system sizes. We see that the relaxation time ($\tau_1$) is size independent  for $L> 200$, which shows that the life time of the
glassy states remains finite in the thermodynamic limit.

\subsection{Boundary conditions}

\begin{figure}
\begin{center}
\includegraphics[scale=0.34,angle=-90]{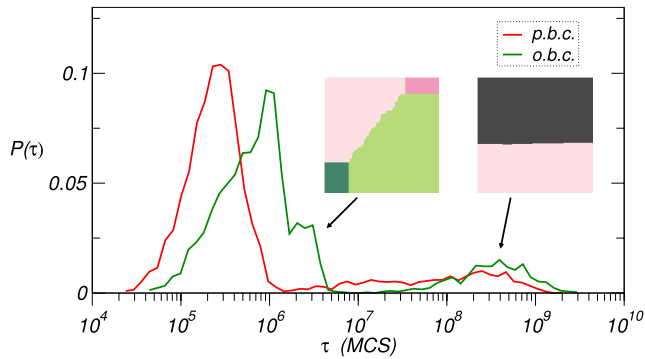}
\caption{\label{fig16} (Color on-line) Equilibration time
probability distribution $P(\tau)$  for $q=9$, $L=200$ and $T=0.2$, using periodic (p.b.c) and open (o.b.c) boundary conditions.
The inset images show typical blocked spin configurations observed with o.p.c.}
\end{center}
\end{figure}

Finally, we analyzed the influence of the boundary conditions in
the relaxation. To this end, we repeated some of the previous
calculations using open boundary conditions. We found that the
overall relaxation scenario found using periodic boundary
conditions repeats qualitatively for open ones. Moreover, the relaxation
time associated with striped configurations appears to be of the same order of magnitude of
that corresponding to periodic boundary conditions. Although a more systematic study should be done to confirm
that, it seems reasonable since the basic activated mechanisms here described should
be still dominant in the case of open boundary conditions. In Fig.\ref{fig16}
we show an example of the equilibration time probability
distribution for $q=9$ and some typical blocked spin
configurations. In this case, Lifshitz states are no longer
composed only by hexagons for relatively small system sizes (due to the presence of the
borders), but we see clearly the presence of stable three-colored
vertex. Indeed, the observed configurations strongly resemble the
blinking states reported in Ref.\cite{SpKrRe2001b}.

\section{Summary and conclusions}

\label{discussion}

\begin{figure}
\vspace*{13pt}
\begin{center}
\includegraphics[scale=0.3,angle=-90]{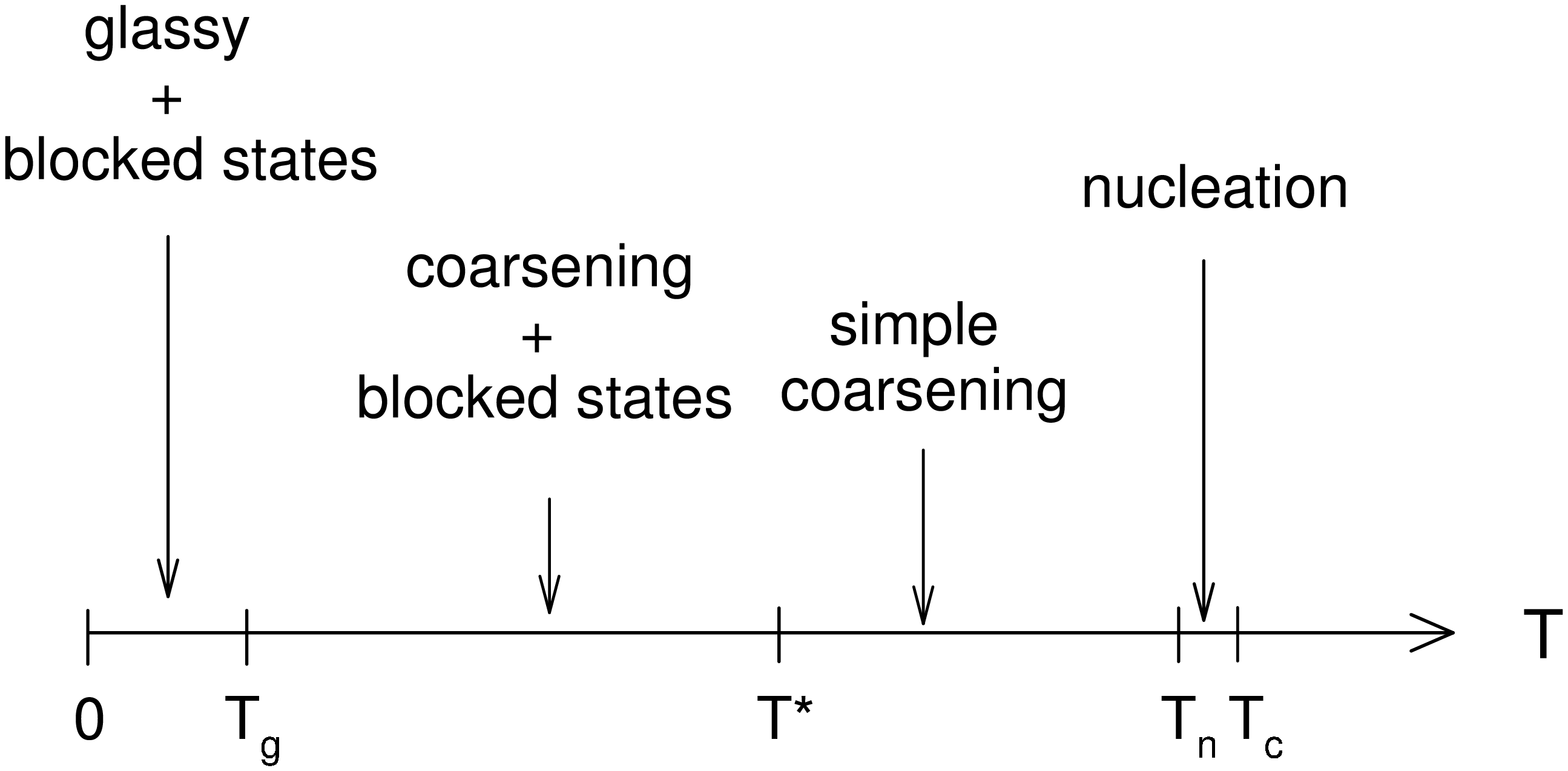}
\caption{\label{fig17} Dynamical regimes in the long term relaxation
of the q-state Potts model with $q>4$, after a quench from
infinite temperatures down to a sub--critical temperature $T$.}
\end{center}
\end{figure}

The main conclusions of this work are summarized in the scheme of
Fig.\ref{fig17}. After a quench from infinite temperature down to
subcritical temperatures, the Potts model with single spin flip kinetics
and periodic boundary conditions presents for $q > 4$
different relaxational regimes, determined by different crossover
characteristic temperatures. Close enough to the critical
temperature, i.e., for $ T_n < T < T_c$, relaxation is dominated
by nucleation mechanisms. For intermediate temperatures $T^* <T <
T_n$ the system crosses over into a simple coarsening dominated
regime where LAC law $l(t) \sim t^{1/2}$ holds until full
equilibration, for most of the realizations of the stochastic
noise. For lower temperatures $T_g<T<T^*$ the normal coarsening
process is interrupted when the system gets stuck  into highly symmetric blocked configurations, composed by
macroscopic ferromagnetic domains, namely, striped and Lifshitz
states. In those cases, the dynamic becomes activated with
characteristic energy barriers, which give rise to distinct time
scales for the different process.

Concerning the role of temperature in the relaxation through blocked states, we found that it has a double
effect: at short time scales it enhances the probability of
reaching them (which is zero at $T=0$) and at long time scales it
allows to escape from them through activation.
At least for $q=9$, our
simulations (for system sizes up to $L=500$) suggest that the
probability of reaching a blocked state at finite temperatures remains
finite when $L\rightarrow\infty$.

Striped states were previously found and characterized for the
Ising model ($q=2$) at very low temperatures. We found that their influence in the
relaxation process is relevant for any value of $q$ , even at
relatively large values of $T$, but their occurrence probability is smaller for $q \geq 3$ than in the Ising case.

We found that the relaxation times associated with the blocked
states present in general the finite size scaling behavior
$\tau\sim L^\omega$, where the exponent $\omega$ depends on $T$, taking values between $2$ and $4$. Such values of the
exponent make the associated time scales several orders of
magnitude larger than those associated with a normal coarsening
process (which scale as $L^2$) for large enough sizes, even at relatively large values of the temperature.

Lifshitz prediction have been recently verified in the phase separation dynamics of diblock copolymers (Cahn-Hillard
model),
in a 2D hexagonal substrate\cite{GoVaVe2006}. We verified that Lifshitz prediction also holds for the $q$ state Potts
model with $q \geq 3$, even in a square lattice, if the system size is large enough.  This strong finite size effects is
 probably due to
the square symmetry of the lattice (large system sizes are required in
order to the influence of the lattice to be faded out) and one
should expect to be reduced in a lattice with three-fold symmetry
(for instance, triangular).

At very low temperatures $T<T_g$
the system gets always trapped in glassy like metastable configurations whose
life time is size--independent and diverge for
$T\rightarrow\infty$. After relaxation from the glassy state, the
system can again gets trapped in a blocked state. Even when the glassy states
do not dominate the relaxation at long enough time scales, a
complete description of the relaxation dynamics cannot exclude
their existence and therefore they deserve further investigations.

Finally, we verified that the whole qualitative relaxation
scenario appears both for periodic and open boundary conditions, although the finite size scaling of the relaxation times may differ in both cases.

Fruitful discussions with M. Iba\~nez de Berganza, F. A. Tamarit and C. B. Budde are
acknowledged. This work was partially supported by grants from
CONICET (Argentina), SeCyT, Universidad Nacional de C\'ordoba
(Argentina), FONCyT grant PICT-2005 33305 (Argentina) and ICTP
grant NET-61 (Italy).

\end{document}